\documentclass[%
 aip,
 amsmath,amssymb,
 reprint,%
]{revtex4-1}

\usepackage{amsmath}
\usepackage{graphicx}
\usepackage{dcolumn}
\usepackage{bm}
\usepackage{xcolor}

\usepackage[utf8]{inputenc}
\usepackage[T1]{fontenc}
\usepackage{mathptmx}
\usepackage{etoolbox}
\usepackage{cleveref}
\Crefformat{figure}{#2Fig.~#1#3}
\Crefmultiformat{figure}{Figs.~#2#1#3}{ and~#2#1#3}{, #2#1#3}{ and~#2#1#3}
\usepackage{subfig}

\makeatletter
\def\@email#1#2{%
 \endgroup
 \patchcmd{\titleblock@produce}
  {\frontmatter@RRAPformat}
  {\frontmatter@RRAPformat{\produce@RRAP{*#1\href{mailto:#2}{#2}}}\frontmatter@RRAPformat}
  {}{}
}%
\makeatother
\begin{document}

\preprint{AIP/123-QED}

\title[]{Kinetic phase diagram for two-step nucleation in colloid-polymer mixtures \\~}

\author{Willem Gispen}
 \affiliation{ 
Soft Condensed Matter \& Biophysics, Debye Institute for Nanomaterials Science, Utrecht University, Princetonplein 1, 3584 CC Utrecht, the Netherlands 
}%
\author{Peter G. Bolhuis}
\affiliation{ 
van 't Hoff Institute for Molecular Sciences, University of Amsterdam, P.O. Box 94157, 1090 GD Amsterdam, The Netherlands
}%
\author{Marjolein Dijkstra}%
 \email{m.dijkstra@uu.nl}
 \email{w.h.gispen2@uu.nl}
\affiliation{ 
Soft Condensed Matter \& Biophysics, Debye Institute for Nanomaterials Science, Utrecht University, Princetonplein 1, 3584 CC Utrecht, the Netherlands 
}%

\date{\today}

\begin{abstract}
\vspace{1em}
{\sffamily \textbf{ABSTRACT}}\\ ~\\ 
Two-step crystallization via a metastable intermediate phase is often regarded as a non-classical process that lies beyond the framework of classical nucleation theory (CNT).
In this work, we investigate two-step crystallization in colloid-polymer mixtures via an intermediate liquid phase.
Using CNT-based seeding simulations, we construct a kinetic phase diagram that 
identifies regions of phase space where the critical nucleus is either liquid or crystalline.
These predictions are validated using transition path sampling simulations at nine different relevant state points.
When the critical nucleus is liquid, crystallization occurs stochastically during the growth phase, whereas for a  crystalline critical nucleus, the crystallization process happens pre-critically at a fixed nucleus size. We conclude that CNT-based kinetic phase diagrams are a powerful tool for understanding and predicting `non-classical' crystal nucleation  mechanisms. 
\vspace{3em}
\end{abstract}

\maketitle

\section{Introduction}

Crystal nucleation can be facilitated by the formation of a metastable intermediate phase, a phenomenon commonly known  as two-step nucleation.
This mechanism can increase the crystal nucleation rate by several orders of magnitude,\cite{wolde_enhancement_1997, wedekind_optimization_2015} making it  undoubtedly relevant for  understanding and predicting  crystallization in both natural and industrial processes.
Two-step nucleation is often  described as `non-classical' because the stable crystal phase does not form directly.\cite{gvekilov_two-step_2010, sear_non-classical_2012, sosso_crystal_2016, lee_nonclassical_2016, karthika_review_2016, blow_seven_2021, jun_classical_2022, finney_molecular_2024}
This non-classical view necessitates extending  classical nucleation theory (CNT), such as modifying the free-energy landscape, \cite{iwamatsu_free-energy_2011,eaton_free_2021,bulutoglu_investigation_2022,bowles_influence_2022} or accounting for anisotropic kinetic effects.\cite{addula_molecular_2021}
In doing so, it is easy to overlook that standard CNT still remains a powerful tool for understanding  and predicting nucleation processes.
For example, the CNT-based seeding approach\cite{espinosa_seeding_2016} efficiently approximates homogeneous nucleation barriers and rates using computer simulations.
The seeding approach has been demonstrated to capture nucleation trends for a wide range of systems, thermodynamic conditions, and particle interactions.\cite{espinosa_seeding_2016, espinosa_interfacial_2016, gispen_brute-force_2023, grabowska_homogeneous_2023, gispen_finding_2024}
It is also possible to construct  `kinetic phase diagrams', providing insights into the nucleation and growth of metastable crystal polymorphs.\cite{sadigh_metastablesolid_2021, gispen_kinetic_2022} 

However, during homogeneous nucleation from a gas, solution, or dilute suspension, the intermediate phase is typically a disordered phase, such as a liquid or amorphous solid, rather than a  metastable crystal polymorph.
This two-step mechanism has been observed in various complex systems  
such as methane hydrates,\cite{arjun_unbiased_2019} zeolites, \cite{kumar2018could,kumar2018two,dhabal2021coarse}  NaCl,\cite{bulutoglu_investigation_2022} insulin,\cite{kaissaratos_two-step_2021} lysozyme,\cite{gvekilov_two-step_2010} 
Lennard-Jones,\cite{van_meel_two-step_2008,chen_aggregation-volume-bias_2008,lutsko_how_2019} lattice models,\cite{hedges_limit_2011, eaton_free_2021,duff_nucleation_2009} and colloidal systems.\cite{fortini_crystallization_2008,khaxton_crystallization_2015,poon_kinetics_2000,renth_phase_2001,huizhang_can_2019,savage_experimental_2009}
Although two-step nucleation via a disordered phase is observed in a wide range of systems and significantly impacts  the crystal nucleation rate, it has never been investigated whether CNT-based seeding simulations can be successfully applied to this `non-classical' process.

We investigate this question using  computer simulations of a model for colloid-polymer mixtures. This system is chosen because the effective attraction strength and range between the colloids can be easily tuned, yielding rich phase behavior.
For example, brute-force simulations of colloid-polymer mixtures have revealed regions of phase space where different phase transition phenomena occur, including one-step and two-step nucleation, glass or gel formation, spinodal decomposition, and the formation of metastable liquid phases.\cite{fortini_crystallization_2008, khaxton_crystallization_2015}
Furthermore, the larger time and length scales of colloids  facilitate experimental observations of the nucleation process.


In this work, we employ extensive seeding\cite{espinosa_seeding_2016} and transition path sampling\cite{bolhuis_transition_2002} simulations to study crystal nucleation from a dilute suspension of colloids with polymer-induced attractions.
Using seeding simulations and classical nucleation theory,  we predict a kinetic phase diagram of nucleation. 
This kinetic phase diagram describes  regions of phase space where the critical nucleus is either crystalline or liquid, roughly corresponding to the regions of one-step and two-step crystallization.
Of course, these predictions need to be validated with independent, unbiased methods.
For this purpose, we use transition path sampling to capture the nucleation mechanism at nine different state points around the boundary of the above-mentioned regions.
Our findings indicate that CNT provides  valid predictions of crystallization mechanisms even in the case of 'non-classical' two-step nucleation.

\section{Model system}
We consider a model system of hard-core colloids and polymer depletants introduced independently by Asakura and Oosawa\cite{asakura_interaction_1954} and by Vrij.\cite{vrij_polymers_1976}
In such a system, the polymers are considered as ideal chains that  induce an entropic attractive depletion interaction between the colloids, which we model with an effective Asakura-Oosawa-Vrij (AOV) pair potential between the colloids.\cite{dijkstra_phase_1999}
The pair potential is determined by the polymer reservoir packing fraction $\phi$ and the size ratio $q=\sigma_p / \sigma$ between the polymer diameter $\sigma_p$ and the colloid diameter $\sigma$. In the pair-potential approximation, the effective attractive interaction energy $u_{\mathrm{AOV}}(r)$ between a pair of colloids at a distance $r/\sigma < 1+q$ is given by
\begin{eqnarray}
    \label{eq:ao}
    \beta u_{\mathrm{AOV}}(r)&=& \nonumber \\ 
&& \hspace{-20mm} -\phi \frac{(1+q)^3}{q^3} \left [ 1 - \frac{3}{2(1+q)} \frac{r}{\sigma} + \frac{1}{2(1+q)^3} \left(\frac{r}{\sigma}\right)^3 \right],
\end{eqnarray}
where $\beta = 1/(k_B T)$ represents the inverse thermal energy. As the polymers are considered as an ideal gas, no interactions are present between colloids at distances $r/\sigma \geq 1+q$.
By using this pair potential approximation, we reduce the computational complexity of our simulations by avoiding the explicit simulation of polymers.
We focus on the case $q=0.6$ and vary the polymer reservoir packing fraction $\phi$ from $0$ to $1$. 
Note that the attraction strength is directly proportional to $\phi$, with the maximum potential well depth considered being $3.3 ~k_B T$ at $\phi=1$.
To model the hard-core interaction between the colloids, we use the pseudo-hard-sphere potential $\beta u_{\mathrm{pHS}}(r)$ introduced in Ref.\ \citenum{jover_pseudo_2012}. This is a Weeks-Chandler-Andersen (WCA)-like potential that acts as a steeply repulsive potential for particles at distances  $r/\sigma < 1.02$.
By using a higher set of exponents ($49$ and $50$) than a standard WCA potential ($6$ and $12$), the pseudo-hard-sphere potential is more steeply repulsive than a standard WCA potential and closely mimics the hard-sphere equation of state, such that rescaling with an effective hard-sphere diameter is not necessary.\cite{jover_pseudo_2012}
The total pair potential $\beta u(r) = \beta u_{\mathrm{AOV}}(r) + \beta u_{\mathrm{pHS}}(r)$ is simply given by the sum of the Asakura-Oosawa-Vrij and the pseudo-hard-sphere potential. The minimum of the total potential is also located at $r/\sigma \approx 1.02$.
We perform molecular dynamics and Langevin dynamics simulations of particles interacting with this pair potential $u(r)$.
The calculations for the equilibrium phase diagram are performed using  molecular dynamics simulations in the isobaric-isothermal ($NPT$) ensemble, while seeding and transition path sampling simulations are carried out using Langevin dynamics in the grand-canonical $(\mu VT)$ ensemble.  See \Cref{sec:simulation-details} for more simulation details.

\begin{figure}[!t]
    \centering
    \subfloat{
    \includegraphics[width=\linewidth]{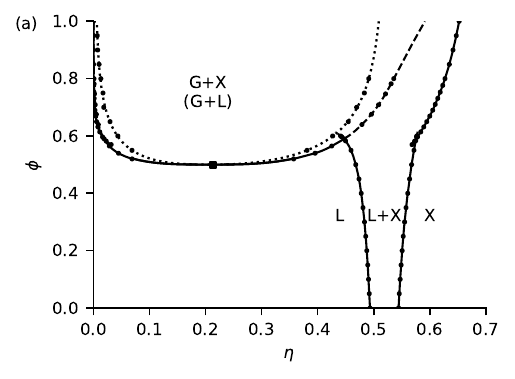}
    }
    \\
    \subfloat{
    \includegraphics[width=\linewidth, ]{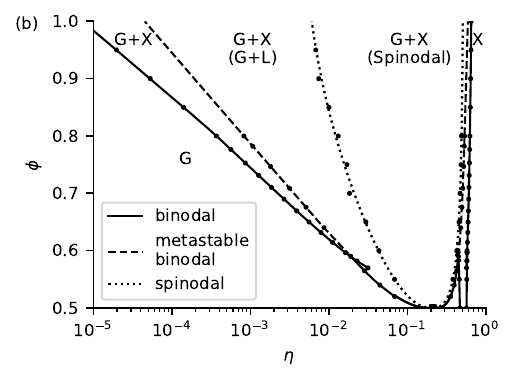}
    }
    \caption{Equilibrium phase diagram of a colloid-polymer mixture with a size ratio of $q=\sigma_p/\sigma=0.6$, where $\sigma_p$ represents the polymer diameter and $\sigma$ the colloid diameter, shown in the colloid packing fraction $\eta$ -- polymer reservoir packing fraction $\phi$ plane, with $\eta$ presented on a linear (a) and on a logarithmic scale (b). Solid lines represent stable binodals, dashed lines indicate metastable binodals, dotted lines correspond to spinodals. The labels G, L, X denote the gas, liquid, and crystalline solid phases, respectively.}
    \label{fig:eq-phase-diagram}
\end{figure}

\section{Equilibrium phase diagram}
We first determined the equilibrium phase diagram using free-energy calculations and Gibbs-Duhem integration of the binodal lines (see \Cref{sec:eq-phase-diagram-details} for more details). In \Cref{fig:eq-phase-diagram}, we show the equilibrium phase diagram in the $(\eta, \phi)$-plane, where $\eta$ is the colloid packing fraction and $\phi$ is the polymer reservoir packing fraction. 
Above the critical point, which is marked with a black square at $\eta=0.21$ and $\phi=0.50$, one can distinguish between a dilute colloidal fluid and a dense colloidal fluid. We will  refer to these phases as the gas and liquid phases, respectively, in analogy to molecular systems. The stable binodals, metastable binodals, and spinodals are represented by solid, dashed, and dotted lines, respectively. The triple point is located at $\phi=0.59$ with a liquid phase density of $\eta=0.45$, in good agreement with Ref.\ \citenum{dijkstra_phase_1999}. The letters G, L, and X refer to the regions where the gas (G), liquid (L), and crystalline solid (X) phases are stable. We have also marked the coexistence regions, for example L+X indicates the region where a liquid-crystal coexistence is stable.


To reach our goal of studying the effectiveness of CNT for two-step nucleation processes we investigate  
crystal nucleation from the supersaturated gas phase.
Using the equilibrium phase diagram, we can already identify the regimes where nucleation from the gas phase is possible.
In \Cref{fig:eq-phase-diagram}(b), we have plotted $\eta$ on a logarithmic scale, with the focus on $\phi>0.5$,  to highlight the low-packing-fraction region above the critical point. At very low colloid packing fractions $\eta$, the region marked `G' corresponds to the stable gas phase, where nucleation is impossible.
As we increase $\eta$, we cross the gas-crystal binodal and enter a region labeled `G+X', where the gas phase is metastable with respect to a gas-crystal coexistence and the gas-liquid coexistence is unstable.
Therefore, nucleation of the crystal phase is possible in this regime.
At even higher $\eta$, we cross the gas-liquid binodal and reach the region labeled `G+X (G+L)', where the gas phase is metastable with respect to both gas-crystal  and  gas-liquid coexistence (indicated by the round brackets). 
Here, the liquid state is metastable and the system thermodynamically prefers a gas-crystal coexistence. 
Finally, crossing the dotted spinodal line, we enter  the spinodal regime, where the gas phase is unstable with respect to  spinodal decomposition.

From now on, we will focus on the regions `G+X' and `G+X (G+L)' between the gas-crystal binodal and the gas spinodal.
According to the equilibrium phase diagram, the gas phase in these regions is metastable and will eventually relax into a gas-crystal coexistence.
However,  the nucleation mechanism in these regions is  unclear, as is the role of  a two-step nucleation process.

\begin{figure*}[!t]
    \centering
    \includegraphics[width=0.8\linewidth]{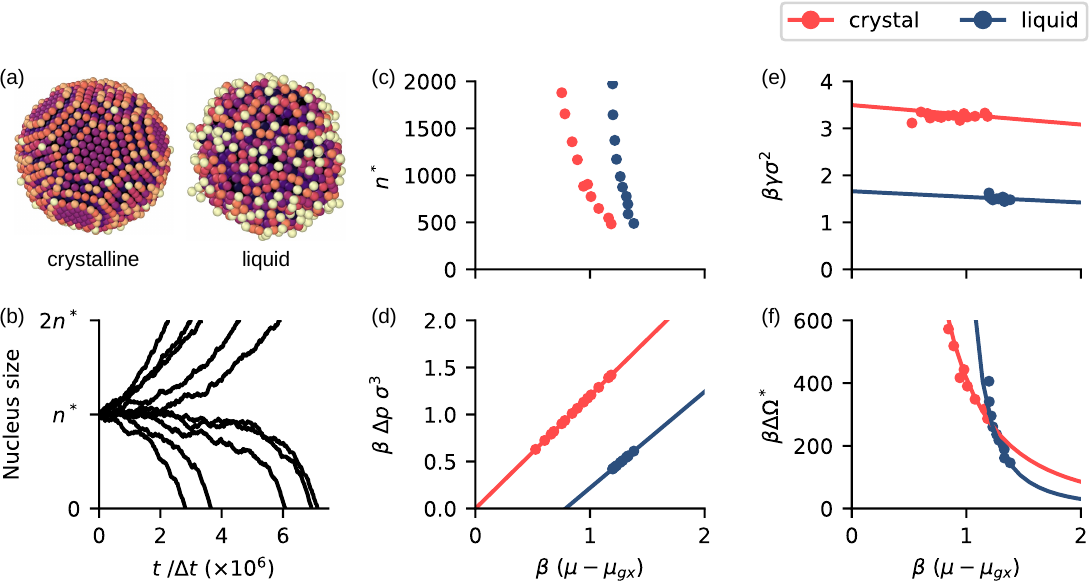}
    \caption{Illustration of the seeding method at a polymer reservoir packing fraction of $\phi=0.8$. (a) Crystal (left) and liquid (right) seeds are inserted into a metastable gas phase. The external surfaces of the seeds are shown, and particles are shown in a darker color if they have more nearest neighbours. Particles in the surrounding gas phase are not shown for visual clarity. (b) For each seed, the chemical potential is identified at which the seed is critical, i.e. where it grows and shrinks with equal probability. To this end, we plot the nucleus size $n$ as a function of time measured in units of  $\Delta t = 0.001 \sqrt{\beta m \sigma^2}$ for ten independent simulation trajectories. 
    Using the critical nucleus size $n^*$ (c) and the pressure difference $\Delta p$ (d), we  calculate the interfacial tension $\gamma$ (e) and nucleation barrier $\Delta \Omega^*$ (f) as a function of the chemical potential $\mu$, which is  plotted relative to the chemical potential $\mu_{gx}$ at gas-crystal coexistence.}
    \label{fig:seeding}
\end{figure*}

\section{Nucleation barriers from seeding simulations}
To unravel the nucleation mechanism, we perform extensive seeding simulations to calculate the free-energy barriers for nucleation of the liquid and crystal phases from the metastable gas phase.
The seeding technique\cite{espinosa_seeding_2016} relies on classical nucleation theory (CNT) to obtain approximations of the free-energy barriers for nucleation.
To explain the relevant CNT equations, consider a metastable gas (`parent') phase and a `child' phase (liquid or crystal) at the same chemical potential $\mu$ in the grand canonical $(\mu VT)$ ensemble. According to CNT, the change in grand potential (or free-energy difference) $\Delta \Omega$ associated with the formation  of a liquid or crystal nucleus consisting of $n$ particles  from the metastable gas phase is given by
\begin{equation}
    \label{eq:cnt-barrier}
    \Delta \Omega(n) = -v(n) \Delta p + a(n) \gamma.
\end{equation}
Here, $\gamma$ is the interfacial tension and $\Delta p$ is the pressure difference between the bulk gas phase and the child phase at chemical potential $\mu$. 
In the grand-canonical ensemble, $\Delta p$ serves as the driving force for nucleation, similar to $\Delta \mu$ in the isobaric-isothermal ($NPT$) ensemble.
Assuming a spherical shape, the volume $v(n)$, surface area $a(n)$,  and radius $r(n)$ of the nucleus can be computed from the number of particles $n$ in the nucleus and the density $\rho$ of the bulk child phase as
$v(n) = n / \rho$, $r(n) = \left ( {3 n }/{4 \pi \rho} \right )^{1/3}$ and $a(n)=4\pi r(n)^2$.
The height $\Delta \Omega^*$ of the nucleation barrier and the interfacial  tension $\gamma$ are related to the size $n^*$ of the critical nucleus as
\begin{subequations}
\label{eq:seeding}
 \begin{align}
        \Delta \Omega^* &= \frac{1}{2}  v(n^*) \Delta p, \label{eq:seeding-dOmega} \\
        \gamma          &= \frac{1}{2} r(n^*) \Delta p. \label{eq:seeding-gamma}
\end{align}
\end{subequations}
Note that the absolute value of the interfacial tension, as given by \Cref{eq:seeding-gamma}, depends on the spherical-shape assumption. However,  the value of the nucleation barrier, as given by \Cref{eq:seeding-dOmega}, is also valid for any other shape.\cite{gispen_variational_2024, sharma_nucleus-size_2018}
The key point is that we can compute $\Delta \Omega^*$ and $\gamma$ from $\Delta p$ and $n^*$.

The idea of the seeding approach\cite{espinosa_seeding_2016} is to use simulations to identify the critical nucleus size $n^*$ and use  CNT \Cref{eq:seeding-dOmega} to approximate the nucleation barrier.
The seeding approach uses a series of `seeds' - nuclei of varying sizes -  and uses simulations to identify the conditions for which these seeds are critical.
To be more precise, we obtain a spherical seed by cutting a spherical region from a bulk liquid or face-centered cubic (fcc) crystal phase. This seed is then inserted into the metastable gas phase.
Next, we use grand-canonical Langevin dynamics simulations with varying chemical potentials $\mu$ to observe whether this seed grows or shrinks. In this way, we find the chemical potential $\mu^*$ for which the seed grows and shrinks with equal probability. In other words, the seed is critical at this chemical potential. Note that our approach is slightly different from Ref.\ \citenum{espinosa_seeding_2016}, because we vary the chemical potential instead of the pressure or temperature. 

We have illustrated the seeding approach \cite{espinosa_seeding_2016} for a series of crystal and liquid seeds simulated at a polymer reservoir packing fraction $\phi=0.8$ in \Cref{fig:seeding}.
Two exemplary seeds are shown in \Cref{fig:seeding}(a), where the colloidal particles are shown in a darker color if they have more nearest neighbors. 
The left one corresponds to a crystal seed, and the right one to a liquid seed. 
When the seed is critical, it grows and shrinks with equal probability. To illustrate this, we plot the nucleus size $n$ as a function of time for ten independent simulations of the example crystal seed in \Cref{fig:seeding}(b). 
We observe five trajectories where the nucleus grows and five where it shrinks. 
Using a series of crystal and liquid seeds, we obtain $n^*$ as a function of $\mu$. In \Cref{fig:seeding}(c-f), each symbol corresponds to a different seed.
To compute the barrier height, we also need the pressure difference $\Delta p$ shown in \Cref{fig:seeding}(d). $\Delta p$ is obtained from the equations of state $p(\mu)$ of the bulk gas, liquid, and crystal phases, see \Cref{sec:eq-phase-diagram-details} for more details.
The CNT approximations \Cref{eq:seeding} for the interfacial tension $\gamma$ and barrier height $\Delta \Omega^*$ are shown with symbols  in \Cref{fig:seeding}(e,f). \Cref{fig:seeding} illustrates for $\phi=0.8$ how liquid and crystal seeds can be used to obtain the nucleation barriers as a function of the chemical potential.


We have performed extensive seeding simulations for nucleation of both the liquid and  crystal phases with polymer reservoir packing fractions varying between $\phi=0.6$ and $\phi=1.0$. In total, we obtained around $75$ critical crystal seeds and $75$ critical liquid seeds, with sizes ranging from $500$ to $5000$ particles. In other words, for both phases we have an approximation of the critical nucleus size $n^*$ for $75$ different state points $(\mu,\phi)$. 
Using \Cref{eq:seeding}, we can approximate the barrier height and  interfacial tension for each of these state points.
To fit and extrapolate the barrier and interfacial tension as a function of $\mu$ and $\phi$, we first fit the interfacial tension with the fitting function 
\begin{equation}
    \label{eq:gamma-fit}
 \gamma(\mu, \phi) = \gamma_0 \cdot (\phi - \phi_{0})^{\nu} \cdot \left(1 - \gamma_{p} \Delta p(\mu, \phi) \right),
\end{equation}
where $\gamma_0, \phi_0, \nu$, and $\gamma_{p}$ are fitting parameters. We chose this fitting function in order to include a scaling law dependence\cite{brader_fluid-fluid_2000} $(\phi - \phi_{0})^\nu$ on the polymer reservoir packing fraction $\phi$ and a linear dependence\cite{espinosa_seeding_2016} on the pressure difference $\Delta p(\mu, \phi)$.
Next, the barrier height $\Delta\Omega^*$ can be computed from the fitted interfacial tension and the pressure difference $\Delta p$ using the CNT equation
\begin{equation}
     \label{eq:omega-fit}
    \Delta \Omega^*(\mu, \phi) = \frac{16 \pi \gamma(\mu, \phi)^3}{3 |\Delta p(\mu, \phi)|^2},
\end{equation}
which follows from \Cref{eq:seeding}.
We compute the barrier height from the fitted interfacial tension instead of fitting the barrier height directly, because the lower variation in the interfacial tension results in a more stable fit and extrapolation.\cite{espinosa_seeding_2016}
Note that the fits are based on $\phi \geq 0.6$, so any predictions below the triple point at $\phi=0.59$ are based on extrapolations of the fits.
In \Cref{fig:seeding}(e,f), the solid lines represent the fits given by \Cref{eq:gamma-fit} and \Cref{eq:omega-fit} applied to $\phi=0.8$. We see that the fits capture the seeding data well. \Cref{eq:gamma-fit} and \Cref{eq:omega-fit} are fitted separately for the crystal and liquid phases.

Importantly, the fits allow us to approximate $\gamma$ and $\Delta \Omega^*$ for both phases as a function of the chemical potential $\mu$ and the polymer reservoir packing fraction $\phi$.
The key result of our seeding simulations is that we obtain the free-energy barrier height $\Delta \Omega^*$ for  nucleation of both the liquid and  crystal phase across  the entire metastable gas regions `G+X' and `G+X (G+L)' of the phase diagram.

\section{Kinetic phase diagram}    
\begin{figure}[!tbp]
    \centering
    \includegraphics[width=\linewidth]{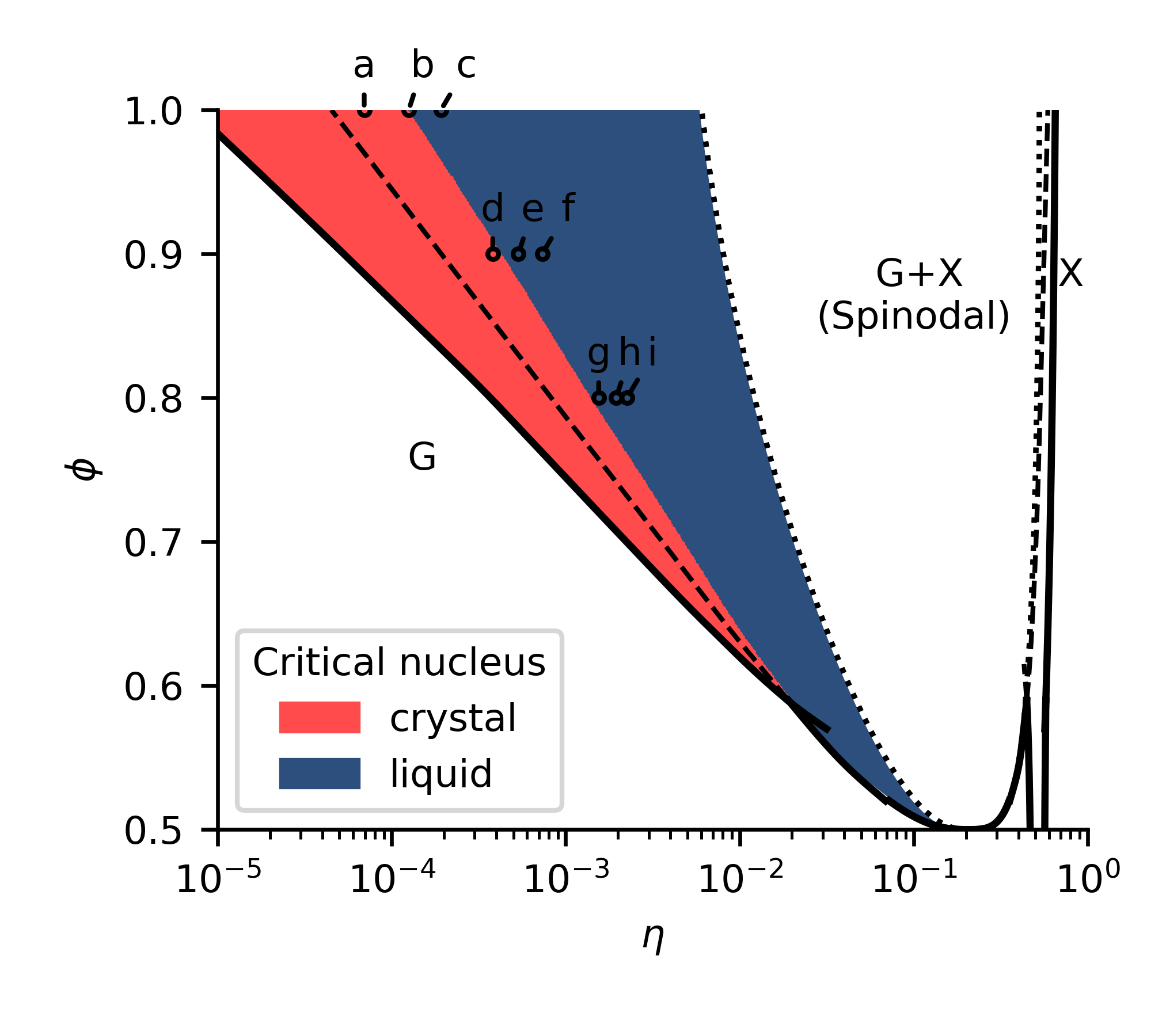}
    \caption{
    Kinetic phase diagram of colloid-polymer mixtures with a size ratio $\sigma_p/\sigma=0.6$ in the colloid packing fraction $\eta$ -- polymer reservoir packing fraction $\phi$ plane. 
    The kinetic phase diagram focuses on  nucleation from the metastable gas phase, denoting the regions where the critical nucleus is crystalline or liquid, as predicted by classical nucleation theory and seeding simulations. 
    The labels (a-i) mark the nine state points examined using transition path sampling, see \Cref{fig:tps-cnt-9}.
    Similarly to \Cref{fig:eq-phase-diagram}, solid lines represent stable binodals, dashed lines denote metastable binodals, and dotted lines indicate spinodals.
    }
    \label{fig:kinetic-phase-diagram}
\end{figure}

We can leverage our knowledge of the nucleation barrier $\Delta \Omega^*$ to construct a `kinetic' phase diagram for nucleation. To be precise, for the entire metastable gas region, we determine the phase that has the lowest nucleation barrier from the gas phase, which   indicates the structure of the most probable critical nucleus. If the liquid phase has a lower nucleation barrier, then the critical nucleus will be (most likely) liquid; if the crystal phase has a lower nucleation barrier,  the critical nucleus will be predominantly crystalline.
In \Cref{fig:kinetic-phase-diagram}, we show the regions where these two scenarios occur in the phase diagram. The blue region indicates where the critical nucleus is liquid, while the red region represents where the critical nucleus is crystalline. Aside from these colored regions, the phase diagram shows the same stable binodals (solid lines), metastable binodals (dashed lines), and spinodals (dotted lines) as in \Cref{fig:eq-phase-diagram}(b).
For low colloid packing fraction $\eta$, the dominant mechanism is the formation of  critical crystal nuclei.
As $\eta$ increases, the primary mechanism shifts to the formation of liquid critical nuclei. The boundary dividing  these two kinetic regimes intersects the triple point and gradually diverges from the gas-liquid binodal. Below the triple point, only liquid critical nuclei are formed.

Both above and below the triple point at $\phi=0.59$, there is a large region where the critical nucleus is liquid. 
This indicates that a metastable liquid phase rather than a  stable crystal phase nucleates from the gas phase in the blue region above the triple point.
The formation of a metastable liquid phase occurs because the liquid-gas interfacial tension is  lower than the crystal-gas interfacial tension.
Since the liquid phase is metastable with respect to the crystal  phase above the triple point, the liquid nuclei will grow and eventually crystallize.
To assess the stability of the liquid phase above the triple point, we measured the crystal nucleation rate from the liquid phase using brute-force molecular dynamics simulations, see \Cref{app:bulk-vs-surface} for more details.
We performed these simulations on bulk liquids at gas-liquid coexistence conditions because the density of the metastable liquid that nucleates from the gas phase will be very close to that of the liquid phase at  gas-liquid coexistence.
From the triple point at $\phi=0.59$ up to $\phi\approx 0.75$, the crystallization rate is very low, and spontaneous crystallization is not observed from a bulk liquid.
Therefore, if a critical liquid nucleus forms in a metastable gas for $\phi \lessapprox 0.75$, this leads to a metastable liquid phase that does not crystallize within the time scale of our brute-force simulations.
In contrast, for $\phi \gtrapprox 0.75$, we observe spontaneous crystallization of bulk liquids. Furthermore, we find that the crystallization rate increases exponentially with the polymer reservoir packing fraction $\phi$. 
Consequently, if a critical liquid nucleus forms in a metastable gas for $\phi \gtrapprox 0.75$, this leads to a growing liquid droplet that will crystallize within a reasonable time scale.
In short, as $\phi$ increases, the liquid phase is further from the triple point and becomes less stable with respect to crystallization.
In other words, the driving force for crystal nucleation increases with $\phi$ along the metastable gas-liquid binodal.

\begin{figure*}[!htbp]
    \centering
    \includegraphics[width=\linewidth]{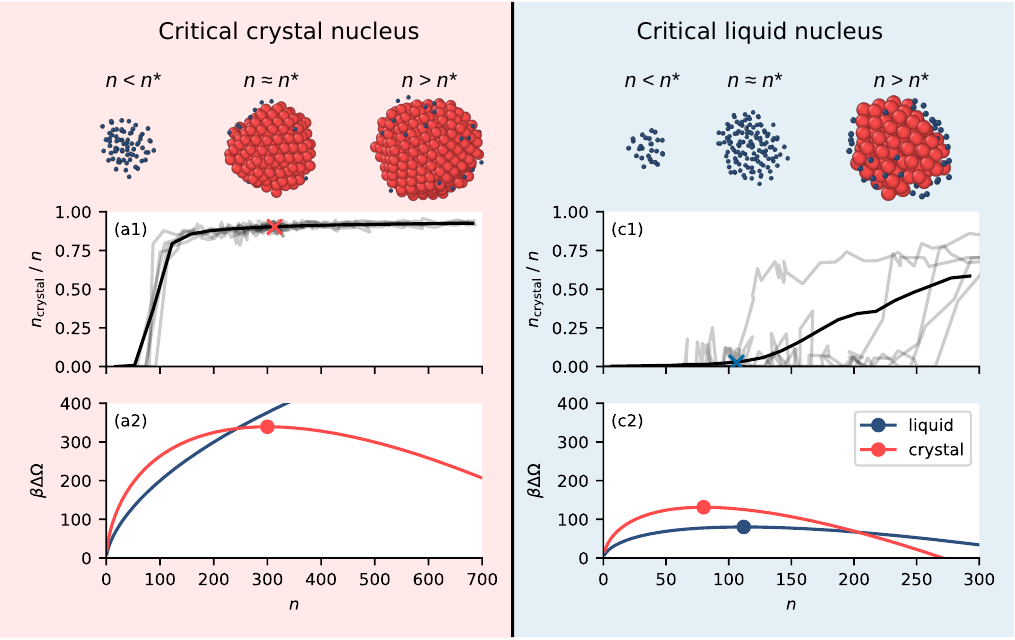}
    \caption{
    Two mechanisms for two-step nucleation corresponding to a critical crystal nucleus (state point (a), left) and critical liquid nucleus (state point (c), right).
    The mechanisms are illustrated here for two (a,c) of the nine state points investigated using transition path sampling.
    At the top, snapshots are shown of pre-critical, critical, and post-critical nuclei. 
    Liquidlike particles are depicted in dark blue and reduced in size, while crystalline particles are shown in red.
    In the middle (a1, c1), the crystallinity, defined as the fraction of crystalline particles $n_{\mathrm{crystal}} / n$, is plotted as a function of the nucleus size $n$.
    The average crystallinity as a function of the nucleus size is represented by a black line, while the average crystallinity and average size of the critical nuclei are indicated by  red or blue crosses.
    Five different transition pathways are shown in light gray to illustrate their variation.
    At the bottom (a2, c2), the free-energy barrier $\beta \Delta \Omega$ for the formation of  crystal or liquid nuclei is shown.
    The nucleation barriers  correspond to the predictions from classical nucleation theory (CNT), where the gas-liquid and gas-crystal interfacial  tensions are measured using seeding simulations. 
}
    \label{fig:tps-cnt-2}
\end{figure*}

\section{Nucleation mechanism from transition path sampling}
Our kinetic phase diagram predicts the regions where the critical nucleus is crystalline or liquid.
However, it is essential to validate these CNT-based predictions through  independent tests.
Additionally, the kinetic phase diagram does not provide   information about the entire formation mechanism, just the phase of the critical nucleus.
Most intriguingly, if a critical nucleus is crystalline, does that imply that the pre-critical nuclei are also crystalline?

The boundary between the two kinetic regimes in the kinetic phase diagram is close to the gas-crystal and gas-liquid binodals. Therefore, the free-energy barrier for nucleation from the metastable gas phase is high near this dividing boundary.
Given the high nucleation barriers involved, it is infeasible to validate the CNT-based predictions and study the nucleation mechanism with brute-force simulations. 
Instead, we perform transition path sampling simulations to obtain unbiased insights into the nucleation mechanism.
Transition path sampling is an enhanced sampling technique specifically designed to efficiently sample transition pathways of rare events like nucleation.\cite{bolhuis_transition_2002} In contrast to other enhanced sampling methods such as umbrella sampling, metadynamics, and forward-flux sampling, transition path sampling is independent of the choice of order parameter used to measure the progress of a transition.\cite{bolhuis_transition_2002} For example, in metadynamics simulations, the selected order parameter directly biases the particle dynamics. This choice of order parameter has been shown to be particularly crucial for two-step nucleation processes.\cite{agarwal_solute_2014, blow_seven_2021} 
For transition path sampling, the only requirement is that the order parameter distinguishes the initial state from the final state, in our case the gas phase from the condensed phases.
As long as this is satisfied, the system is free to choose the nucleation mechanism that is most probable. 
Therefore, to eliminate the influence of the order parameter on our results, we choose to use transition path sampling. Generally speaking, this method generates an ensemble of transition pathways through Monte Carlo `shooting' moves, where each  trial shooting move generates a new pathway based on a previous pathway.\cite{bolhuis_transition_2002}
For each shooting move, a single frame from the previous pathway is selected as a `shooting point'. From the shooting point, the equations of motion are integrated forward and backward in time, thus generating a new pathway.
If the new pathway remains a valid transition pathway,  the move is accepted and the new pathway replaces the previous one, otherwise the previous path is kept. 
In this way, the sampled path ensemble  gradually converges towards the correct nucleation mechanism.

We select nine $(\phi,\eta)$ state points in our kinetic phase diagram that are near the boundary between the two kinetic regimes. These state points are indicated by black circles in \Cref{fig:kinetic-phase-diagram} and correspond to three different polymer reservoir packing fractions $\phi=0.8, 0.9, 1.0$. For each $\phi$, we investigate three different chemical potentials corresponding to the critical chemical potentials of three liquid seeds. The liquid seeds have nucleus sizes of  $n^*\approx 100$ (state points labeled (c,f,i)) , $n^*\approx 200$ (b,e,h) and  $n^*\approx 500$ (a,d,g).
We initialize the transition path sampling simulations with the seeded liquid nuclei.
In this way, it is easy to generate initial transition pathways at the selected state points, and since they originate  from the liquid seeds, these initial transition pathways lead to the formation of a liquid nucleus.
Starting from these initial transition pathways, we generate new transition pathways using transition path sampling with the aimless shooting algorithm\cite{peters_obtaining_2006, mullen_easy_2015} (see \Cref{sec:tps} for further details). The result of our transition path sampling simulations is a converged ensemble of $20-100$ decorrelated nucleation pathways for each state point.
Additionally, due to the nature of the aimless shooting algorithm, the accepted shooting points serve as a proxy for the critical nuclei.
Therefore, we also obtain an ensemble of critical nuclei for each state point.
Next, we use polyhedral template matching\cite{larsen_robust_2016} to quantify the number of crystalline particles $n_{\mathrm{crystal}}$ in each nucleus in all nucleation pathways (see \Cref{sec:ptm} for further details).

Our analysis reveals two distinct nucleation mechanisms corresponding to the two regimes of the kinetic phase diagram: one nucleation mechanism corresponding to a critical crystal nucleus and the other corresponding to a critical liquid nucleus.
In \Cref{fig:tps-cnt-2}, we illustrate the two nucleation mechanisms by focusing on two state points in the kinetic phase diagram: state points (a) and (c).
For each mechanism, we show representative snapshots of the nuclei, the crystallinity $n_{\mathrm{crystal}}/n$ as a function of the nucleus size $n$, and the nucleation barrier $\Delta \Omega(n)$ for nucleation. 
To be specific, the snapshots show examples of pre-critical, critical, and post-critical nuclei from the transition path sampling simulations.
Liquidlike particles are depicted in dark blue and reduced in size, while crystalline particles are shown in red.
The solid black lines in \Cref{fig:tps-cnt-2}(a1,c1) represent the average crystallinity $n_{\mathrm{crystal}}/n$ of nuclei as a function of size.
We compute the average crystallinity by grouping all configurations from the transition path sampling simulations according to their size.
For each group, we compute the average crystallinity $n_{\mathrm{crystal}}/n$.
Additionally, the red and blue crosses in \Cref{fig:tps-cnt-2}(a1,c1) show the average crystallinity and size of the critical nuclei.
To show the variation of the transition pathways, we also plot five example nucleation pathways with light gray lines.
Figs.~\ref{fig:tps-cnt-2}(a2,c2) show the free-energy barriers $\Delta \Omega$ derived from the CNT \Cref{eq:cnt-barrier}, where the liquid-gas and crystal-gas interfacial tensions $\gamma$ are obtained from seeding simulations fitted using  \Cref{eq:gamma-fit}.
We have also marked  the critical size for a liquid nucleus (blue dot) and the critical size for a crystal nucleus (red dot) in \Cref{fig:tps-cnt-2}(a2,c2).

First, we discuss the nucleation mechanism corresponding to a critical crystal nucleus.
This mechanism occurs when the free energy of a critical crystal nucleus is lower than the free energy of  a critical liquid nucleus, as illustrated in \Cref{fig:tps-cnt-2}(a2).
In \Cref{fig:tps-cnt-2}(a1), we indeed see that the crystallinity $n_{\mathrm{crystal}}/n$ of the critical nucleus (red cross) is high.
Although the critical nucleus is crystalline, the pre-critical nuclei that initially form from the metastable gas remain liquid. This can be seen from the low average crystallinity $n_{\mathrm{crystal}}/n$ of the smallest nuclei in \Cref{fig:tps-cnt-2}(a1). The point at which the average crystallinity reaches around $50\%$  marks the stage  when the pre-critical liquid nuclei crystallize. 
From the five example pathways, we see that most liquid nuclei crystallize at a similar size $n_x \approx 80$.


Second, we discuss the nucleation mechanism corresponding to a critical liquid nucleus.
This mechanism occurs when the free energy of a critical liquid nucleus is lower than that of  a critical crystal nucleus.
Indeed, we see in \Cref{fig:tps-cnt-2}(c2) that  the liquid nucleation barrier (blue dot) is  lower  than  the crystal nucleation barrier (red dot).
In \Cref{fig:tps-cnt-2}(c1), we see that the average crystallinity of the critical nucleus (blue cross) is low, indicating that the critical nucleus is indeed liquid.
The average crystallinity starts to increase gradually during the post-critical growth phase of the nucleus.
From the five example pathways in \Cref{fig:tps-cnt-2}(c1), we see that they all crystallize at different nucleus sizes.

Note that there seems to be an asymmetry between the critical crystal nucleus and critical liquid nucleus mechanisms.  Pre-critical crystallization  occurs fast and roughly at the same size $n_x$, while post-critical crystallization is more stochastic.
The latter can be explained by an additional nucleation barrier for crystallization within  the growing liquid droplet. In the first mechanism instead the system has to form the crystal within the time the nucleus grows to the critical size. Only those trajectories that exhibit this fast growing crystallization process will survive, paths that take too long to form a crystal are simply never becoming critical, or will dissolve again before they can.   


\begin{figure*}[!htp]
    \centering
    \includegraphics[width=\linewidth]{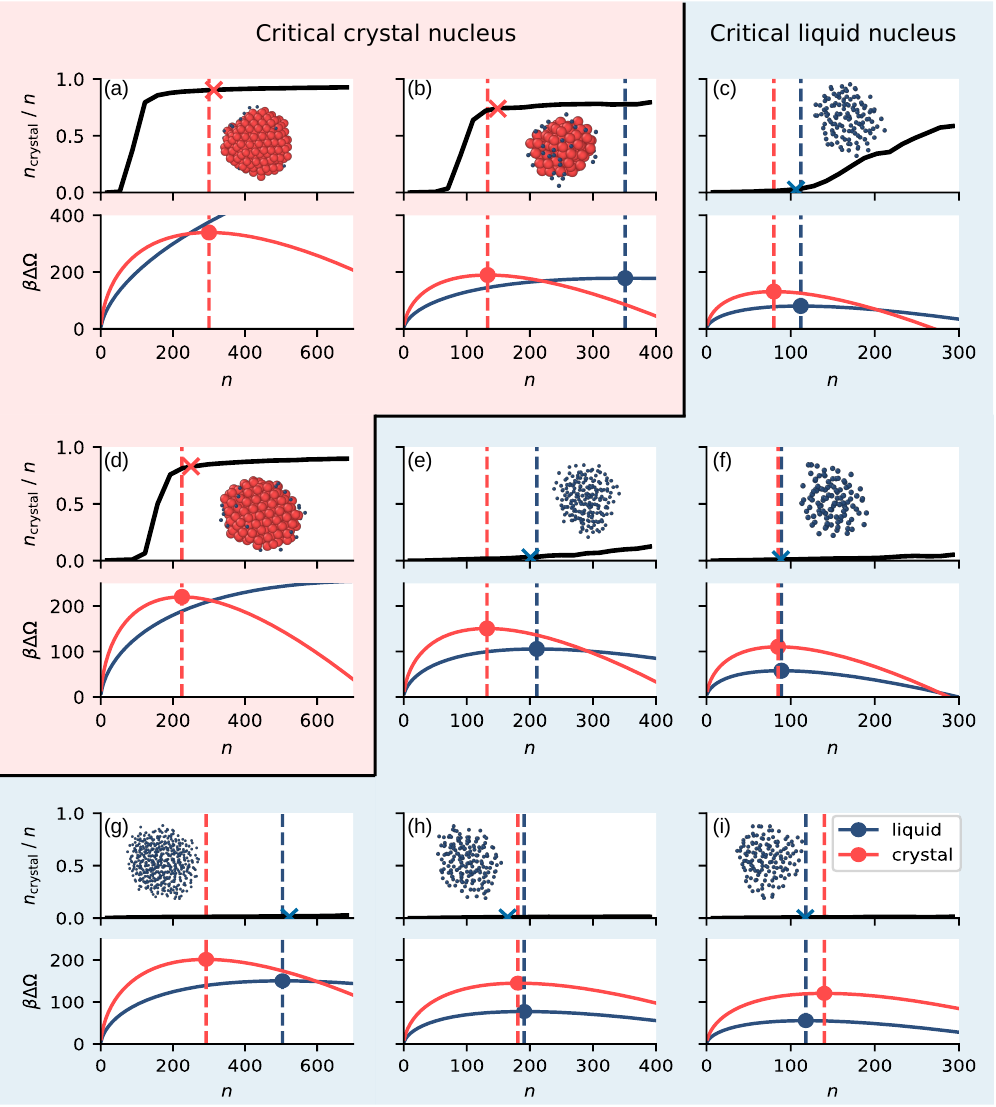}
    \caption{
    Nucleation mechanisms for the nine state points (a-i) investigated using transition path sampling.
    For each state point, a representative snapshot is shown of a critical nucleus.
    Furthermore, for each state point we plot the average crystallinity $n_{\mathrm{crystal}} / n$ as a function of nucleus size $n$ denoted by a solid black line and the critical size represented by a blue or red cross obtained by  transition path sampling 
    and the nucleation barriers $\beta \Delta \Omega$ predicted by classical nucleation theory and seeding.
    The vertical dashed lines and red and blue dots mark the critical nucleus sizes predicted by classical nucleation theory and seeding.
    The state points are divided in two groups: those with a critical crystal nucleus (a,b,d, light red background) and those with a critical liquid nucleus (c,e-i, light blue background).
    Please see the caption of \Cref{fig:tps-cnt-2} for more details.
    }
    \label{fig:tps-cnt-9}
\end{figure*}


Now we have discussed the qualitative features of the two nucleation mechanisms, we present the same information for all nine state points investigated using transition path sampling. In \Cref{fig:tps-cnt-9}, we show snapshots of critical nuclei, average crystallinities and nucleation barriers for each state point (a-i). Again, the average crystallinities are computed from the transition path sampling simulations while the nucleation barriers are from the CNT-based seeding predictions.
Using \Cref{fig:tps-cnt-9} we can evaluate the predictions of our kinetic phase diagram.
The average crystallinity $n_{\mathrm{crystal}}/n$ of the critical nuclei is high for state points (a,b,d) and almost zero for state points (c,e-i).
This indicates that the critical nucleus is crystalline for state points (a,b,d) and liquid for state points (c,e-i), which aligns well with the predictions of the kinetic phase diagram. Notably, state point (b) lies on the boundary between the two kinetic regimes in the kinetic phase diagram. In all other cases, the crystallinity of the critical nucleus clearly supports the predictions of the kinetic phase diagram.
Furthermore, the CNT prediction of the critical nucleus sizes (vertical dashed lines and red and blue dots) compare very well to the critical nucleus sizes measured in transition path sampling (red and blue crosses). 

\section{Discussion}
With our kinetic phase diagram, we have demonstrated that classical nucleation theory (CNT) combined with seeding simulations, can effectively  predict the structure and size of the critical nucleus. In this section, we explore the relationship between equilibrium and kinetic phase diagrams, discuss the limitations of the CNT-based approach, and propose potential extensions  to better capture the phenomenology of two-step nucleation.

\subsection{Relation between equilibrium and kinetic phase diagrams}
We have mapped out the kinetic phase diagram for two-step nucleation in colloid-polymer mixtures in the dilute colloidal gas regime and validated this kinetic phase diagram using transition path sampling simulations. This provides a unique opportunity to compare the kinetic phase diagram (\Cref{fig:kinetic-phase-diagram})  with the equilibrium phase diagram (\Cref{fig:eq-phase-diagram}(b)), allowing us to test theoretical and phenomenological predictions regarding the relationship between equilibrium and kinetic phase diagrams. 

For example, Ostwald's step rule\cite{ostwald_studies_1897} states that the less stable phase will nucleate first.
Stranski and Totomanov\cite{stranski_rate_1933} conjectured instead that the phase that has the lowest nucleation barrier will form first.
Although both Ostwald's step rule and the Stranski-Totomanov conjecture are frequently invoked to explain nucleation phenomena, they are not universally true.
For example, violations of both Ostwald's step rule and the Stranski-Totomanov conjecture have been demonstrated for patchy particles\cite{hedges_limit_2011} and charged colloids.\cite{gispen_kinetic_2022, sanz_evidence_2007}
With our kinetic phase diagram presented in \Cref{fig:kinetic-phase-diagram}, we can compare Ostwald's step rule to the Stranski-Totomanov conjecture.
In the blue region above the triple point, the liquid phase is less stable than the crystal phase, and the critical nucleus is liquid.
Thus, in this region, Ostwald's step rule is obeyed.
In contrast, this rule does not appear to hold in the red region. Here, although the liquid phase is less stable than the crystal phase, the critical nucleus is crystalline. Ostwald's step rule seems to be violated because the barrier to the stable phase is lower than that to the metastable phase.
However, as Ostwald already noted,\cite{ostwald_studies_1897, cardew_ostwald_2023} exceptions to this rule can occur when the transition to the stable phase proceeds so rapidly that the metastable phase is not observed. This likely explains the behavior in this region.
While the critical nucleus is crystalline, the initial pre-critical nuclei that form from the metastable gas phase are liquid. This phenomenon, observed in transition path sampling simulations, aligns with the lower liquid-gas interfacial tension. Such behavior is expected to persist throughout the entire red region.\cite{james_phase_2019}
Ostwald's step rule is also not obeyed in the blue region below the triple point. In principle, there is a metastable crystal-gas coexistence region below the triple point.
The crystal phase is less stable than the liquid phase in this region, but the critical nucleus is liquid. Again, Ostwald's step rule is violated because the barrier to the stable phase is lower than that to the metastable phase.

\citet{evans_classification_2001}\cite{poon_kinetics_2000}\cite{renth_phase_2001} developed a theoretical procedure to predict possible nucleation pathways from the equilibrium phase diagram. In the context of this paper, their procedure assesses whether the driving force for nucleation $\Delta p$ is positive or negative, which can be straightforwardly determined from the (metastable) binodals.
For this discussion, we focus on the two regions above the triple point labeled `G+X' and `G+X (G+L)' in \Cref{fig:eq-phase-diagram}(b). These regions are denoted as
regions `E' and `F', respectively, in Ref. \citenum{evans_classification_2001}.
In the G+X region located left of the gas-liquid binodal, the driving force for nucleation of the liquid phase is negative, while the driving force for nucleation of the crystal phase is positive. Ref. \citenum{evans_classification_2001} predicts that in this region the crystal phase forms directly from the gas phase due to the negative driving force for liquid nucleation. This prediction aligns with our kinetic phase diagram, where the critical nucleus is always crystalline in the G+X region. However, we have observed that small initial pre-critical nuclei can be liquid even if the critical nucleus is crystalline. The critical nucleus size in the G+X region  is very large, which limited our ability to  perform transition path sampling simulations in this region. Nonetheless, small pre-critical liquid nuclei are likely to  play a role here due to the lower gas-liquid interfacial tension.\cite{james_phase_2019}
In the `G+X (G+L)' region   between the gas-liquid binodal and the gas spinodal, both driving forces are positive. Ref. \citenum{evans_classification_2001} predicts that both direct crystal nucleation and a two-step nucleation mechanism via an intermediate  liquid  phase are possible here, but  does not predict which of the  two is favored.
In this paper, we refine this prediction by measuring the interfacial tension in addition to the driving force.
Thus, by measuring the interfacial tension, we determine which of the potential nucleation pathways suggested by Ref.\ \citenum{evans_classification_2001} actually occurs.

\subsection{Crystallization of liquid nuclei}


Can we extract more information from the CNT predictions  beyond the height of the nucleation barrier?
In \Cref{fig:tps-cnt-9}, we observe that the free energy $\Delta \Omega$ of very small liquid nuclei is consistently lower than that  of very small crystal nuclei.
This difference arises because  the liquid-gas interfacial tension is lower than the crystal-gas interfacial tension.
It also explains why the crystallinity of small nuclei remains low in our transition path sampling simulations (\Cref{fig:tps-cnt-9}).
Furthermore, for sufficiently large nuclei, the crystal nucleation barrier eventually crosses the liquid nucleation barrier. In the case of state points (h) and (i), this crossing occurs   at a nucleus size larger than $400$, which is why it is not visible in the plot.
This crossing indicates that  the crystal phase becomes  the thermodynamically most  stable phase at these state points.
These general observations provide further insight into why and where two-step crystallization occurs: the key explanation is that the liquid-gas interfacial tension is lower than the crystal-gas interfacial tension.
However, there may also be conditions where  nuclei are crystalline from the start. This would correspond to a situation where the crystallinity of nuclei is high for all nucleus sizes.
We do not observe such a  `one-step' scenario in our transition path sampling simulations.
From the perspective of CNT, this `one-step' scenario could arise,  for example, if the crystal-gas interfacial tension were lower than the liquid-gas interfacial tension.

The next question is whether  the CNT barriers can  provide a more specific prediction for the nucleus size at which a liquid droplet crystallizes.
Our transition path sampling simulations indicate that the answer to this question is qualitatively different for  critical liquid nuclei and  critical crystal nuclei.
If the critical nucleus is crystalline, our transition path sampling simulations suggest that  crystallization occurs at a specific size $n_x$ that depends on the state point, but is less influenced by  the nucleation pathway.
It is tempting to speculate that this size $n_x$ corresponds to the point where the nucleation barriers cross.\cite{james_phase_2019}
The barrier crossings shown in \Cref{fig:tps-cnt-9} occur at $n_x\approx 240, 220, 310$ for state points (a,b,d), respectively.
Alternatively, \citet{duff_nucleation_2009} suggested that $n_x$
corresponds to the point where the gradients of the nucleation barriers $d\Delta \Omega / dn$ cross.
They argue that these gradients represent the size-dependent chemical potentials of the liquid and crystal nuclei, making freezing favorable once  the gradients cross. 
The gradient crossings are predicted to occur at $n_x \approx 70, 70, 90$ for state points (a,b,d), respectively.
Neither the barrier crossings nor the gradient crossings consistently align with the values $n_x\approx 80$ for state points (a,b) and $n_x\approx 160$ for state point (d) that we obtained from transition path sampling.
However, both the barrier and gradient crossings qualitatively agree with the transition path sampling results, showing  that   $n_x$ decreases with increasing polymer reservoir packing fraction $\phi$ and is less strongly dependent on the supersaturation.

If the critical nucleus is liquid, crystallization in the post-critical regime becomes a rare-event process, occurring on timescales comparable to or longer than those required for the nucleus to grow.  
In this case, while the precise nucleus size at which a droplet crystallizes cannot be precisely predicted, the crystallization rate can still be determined.
In \Cref{app:bulk-vs-surface}, we use brute-force simulations to measure the crystal nucleation rates of both liquid nuclei and bulk liquid phases.
We find that the crystal nucleation rate increases exponentially with the polymer reservoir packing fraction $\phi$.
Additionally, we find that the crystal nucleation rate is not affected by the liquid-gas interface, indicating that  crystal nucleation within liquid droplets is a homogeneous nucleation process initiated in the bulk of the liquid droplet rather than at the liquid-gas interface.
Given the large crystal-gas interfacial tension, it is  expected that crystallization occurs in the bulk of the liquid clusters. 
Once the clusters are fully crystallized, we observe in \Cref{app:bulk-vs-surface}  a peculiar fivefold symmetry, where the colloids are arranged in five tetrahedral `wedges' that together form a pentagonal bipyramid.


\subsection{Limitations of the CNT-based kinetic predictions}
The comparisons in the previous section -- between the nucleation barrier crossings and the size $n_x$ at which the nucleus crystallizes --
hint at some limitations in  using classical nucleation theory and seeding simulations to capture (two-step) nucleation behavior.
One limitation of seeding is that the results are sensitive to the method used to measure the size of the seeds.\cite{zimmermann_nacl_2018,gispen_variational_2024}
As seen in \Cref{eq:seeding}, the values of the interfacial  tensions and nucleation barriers are directly influenced by the seed size $n^*$.
We used the coordination number of each particle to determine whether it is part of the nucleus, see \Cref{sec:seeding-details} for details.
While we have verified that the kinetic phase diagram does not change qualitatively, employing other methods to measure  seed sizes may yield slightly different results.
In the future, this issue could be addressed by employing a recently proposed seeding method.\cite{gispen_variational_2024}
While this new method is not completely independent of how the nucleus size is measured, it is far less sensitive to this choice than standard seeding.
This approach may  also lead to more accurate predictions of the nucleus size $n_x$ at which the liquid droplet crystallizes.
Another possible check on the validity of the seeding simulations is to compare the interfacial tension to direct computations for planar crystal-gas and liquid-gas interfaces.\cite{espinosa_seeding_2016} In the case of the crystal-gas interfacial tension, one of the challenges is that the interfacial tension is sensitive to the crystal orientation, whereas the interfacial tension obtained from seeding is an orientational average. Please see Ref.\ \citenum{pasquale_solid-liquid_2024} for a recent review on the challenges of computing interfacial tensions for crystal-fluid interfaces.

The simple form of the CNT in \Cref{eq:cnt-barrier} for the free-energy barrier allows for the determination of the interfacial  tension and barrier height using only the critical nucleus size. However, modifications to \Cref{eq:cnt-barrier} can 
capture the shape of the nucleation barrier more accurately.\cite{filion_crystal_2010,iwamatsu_free-energy_2011,eaton_free_2021,bulutoglu_investigation_2022,bowles_influence_2022,gispen_variational_2024} These modifications may result in  a free-energy barrier that still depends solely on the nucleus size,\cite{filion_crystal_2010, gispen_variational_2024} or they can be designed to represent a two-dimensional free-energy surface that relies on two parameters, such as the nucleus size and  crystallinity.\cite{iwamatsu_free-energy_2011, eaton_free_2021} Possibly, 
the parameters of these extended theories could  be estimated from seeding simulations.\cite{gispen_variational_2024}
The two-dimensional free-energy surface proposed in Ref.\ \citenum{eaton_free_2021} predicts a composite cluster pathway, wherein the critical nucleus is neither completely liquid nor completely crystalline, but rather a composite cluster of a crystal core completely surrounded by a liquid layer.
We do not observe the composite cluster pathway in our transition path sampling simulations.
Rather, the critical nuclei in our transition path sampling simulations are almost completely liquid or almost completely crystalline. For state point (b), the crystallinity of the critical nucleus is around $74\%$, but the nucleus is not completely surrounded by a liquid layer.
In their lattice model, Ref.\ \citenum{eaton_free_2021} identified the composite cluster pathway at state points that are intermediate between the two extremes of critical crystal and  critical liquid nuclei.
If this behavior  is also the case for our system, the composite pathway would likely  occur only in a small region of phase space, such as between state points (d) and (e).
Alternatively, the composite cluster pathway may become more significant near the triple point.
In this region, the difference in the driving force for nucleation of the crystal or liquid phases reduces, while the difference in interfacial tension remains.
These conditions are more conducive for triple-point wetting, making it more likely that the critical nucleus is a composite cluster.

However, there are two complications for observing a composite critical nucleus near the triple point. 
First, as we see in \Cref{fig:tps-cnt-9}, the nucleus size needed to observe a critical crystal nucleus becomes larger as we approach the triple point. For example, for polymer reservoir packing fraction $\phi=0.8$ a nucleus of size $500$ is not large enough to crystallize.
Therefore, to simulate crystalline or composite critical nuclei near the triple point, we need even larger system sizes. 
$NVT$-seeding\cite{rosales-pelaez_seeding_2020} could be a suitable technique to explore these large composite nuclei in the future.
Second, the crystallization barrier from the metastable liquid  increases as we approach the triple point, creating a barrier to  the formation of both composite clusters and completely crystalline  nuclei.
In our approach, we essentially compare the free energy of a completely crystalline critical nucleus with that of a completely liquid critical nucleus.
If the crystallization barrier from the metastable liquid droplet is high,
the formation of the crystal nucleus is inhibited, even if the free energy of the crystalline critical nucleus is lower than that of the critical liquid nucleus.
These considerations are also relevant for two-step nucleation processes where the intermediate phase is a gel, glass, or amorphous solid rather than a liquid phase.
In such processes, crystallization may be  inaccessible due to kinetic trapping. 
Additionally, CNT is in principle an equilibrium theory.
For example, the driving force $\Delta p$ is based on the pressure difference between two bulk phases in equilibrium at the same chemical potential.
Therefore, special care should be taken when applying CNT or seeding to non-equilibrium states such as gels, glasses, or amorphous solids.

\section{Conclusions}
Two-step crystal nucleation via a metastable intermediate phase is frequently described as a `non-classical' nucleation process. 
Using seeding simulations, we can carefully measure the thermodynamic parameters of classical nucleation theory for the competing phases.
In this way, we have
constructed a kinetic phase diagram for colloid-polymer mixtures that identifies regions of phase space where the critical nucleus is either liquid or crystalline. 
Our predictions were validated using transition path sampling at nine different state points.
These results show that classical nucleation theory, when combined with seeding simulations, can effectively capture the `non-classical' two-step nucleation behavior of colloid-polymer mixtures. 
Our transition path sampling simulations demonstrate that the structure of the critical nucleus has a large impact on the crystallization kinetics.
When the critical nucleus is liquid, crystallization occurs stochastically during the growth of the liquid droplet. In this scenario, the lower interfacial tension of the liquid phase with respect to the gas phase reduces the nucleation barrier, thereby enhancing the nucleation rate.
When the critical nucleus is crystalline, crystallization occurs from a liquid droplet that has not yet reached the critical size. In this case, the nucleus size at which crystallization occurs seems to  depend on the state point, however this size does not vary significantly between different nucleation pathways.

\section*{Supplementary material}
The supplementary material contains the code used to generate and analyze the results of this paper.

\begin{acknowledgments}
M.D. and W.G. acknowledge funding from the European Research Council (ERC) under the European Union's Horizon 2020 research and innovation programme (Grant agreement No. ERC-2019-ADG 884902 SoftML).
\end{acknowledgments}

\section*{Data Availability Statement}
The data supporting the findings of this study are available within the article and its supplementary material.

\appendix

\section{Langevin and molecular dynamics simulations}
\label{sec:simulation-details}
All simulations were performed using the LAMMPS molecular dynamics code.\cite{plimpton_fast_1995}
We employed a timestep $\Delta t=0.001 \sqrt{\beta m \sigma^2}$ to integrate the equations of motion, where $m$ is the particle mass (note that we use reduced units).
Calculations for the equilibrium phase diagram were conducted with molecular dynamics simulations in the isobaric-isothermal ($NPT$) ensemble. The temperature and pressure were kept constant using a Nos{\'e}-Hoover thermostat and barostat with relaxation constants of $100 \Delta t$ and $500 \Delta t$, respectively. 
For the seeding simulations and transition path sampling, we used  hybrid Langevin dynamics / Monte Carlo  simulations in the grand-canonical $(\mu VT)$ ensemble. 
The temperature was controlled by the random forces from  the underdamped Langevin dynamics, with a relaxation constant of $100 \Delta t$. This damping/relaxation constant also determines the friction term of the Langevin equation.
The chemical potential was maintained through grand canonical Monte Carlo particle exchanges with an ideal gas reservoir, where insertions and deletions were attempted with equal probability.\cite{frenkel_understanding_2001}
Specifically,  $10$ exchanges were attempted every $10$ simulation timesteps.
Insertions and deletions were attempted everywhere in the simulation volume.
When successfully inserted, the velocity of a particle is randomly generated from the Maxwell-Boltzmann distribution.

\section{Determination of the equilibrium phase diagram}
\label{sec:eq-phase-diagram-details}

\subsection{Binodals from Gibbs-Duhem integration}
We determine the coexistence pressure $p_{\textrm{coex}}$ as a function of the polymer reservoir packing fraction $\phi$ using Gibbs-Duhem integration.\cite{kofke_direct_1993}
This method relies on the fact that we can measure the derivatives $d\mu/d\phi$ and  $d\mu/dp$ directly within a simulation. To be precise, the two derivatives are given by
\begin{align}
   \frac{d \mu}{d \phi} &= \left \langle \frac{du_{\mathrm{pot}}}{d\phi} \right\rangle \equiv \lambda, \\
    \frac{d\mu}{d p} &= (1/\varrho). \label{eq:dmudp}
\end{align}
Here, $u_{\mathrm{pot}}$ represents the potential energy per particle and $\rho$ denotes the colloid number density. $\lambda$ denotes the average of the derivative $du_{\mathrm{pot}} / d\phi$, which  can  be readily calculated from the pair potential as presented in \Cref{eq:ao} and summed over each pair of particles in the system.
We measure the two derivatives using molecular dynamics simulations in the $NPT$ ensemble with $N=4\times 10^3$ particles and calculate the differences $\Delta \lambda=\lambda_\alpha-\lambda_\beta$ and $\Delta (1/\varrho) = \rho_\alpha^{-1}- \rho_\beta^{-1} $ between the coexisting phases $\alpha$ and $\beta$.
Along the coexistence curves, the pressure and  chemical potential of the coexisting phases should remain equal. Therefore, the gradient of the coexistence curve is given by
$$
\frac{d p_{\textrm coex}}{d \phi} = - \frac{\Delta \lambda}{\Delta (1/\varrho)}.
$$
We integrate this ordinary differential equation using an explicit, third-order Runge-Kutta method with an adaptive step size.\cite{bogacki_32_1989}
The integration of the fluid-crystal binodal starts at $\phi=0$ from the (reduced) pseudo-hard-sphere coexistence pressure  $\beta p_{\textrm coex} \sigma^3=11.65$, as determined in Ref.\ \citenum{espinosa_fluid-solid_2013}.
We start the integration of the gas-crystal binodal from an initial point obtained from an Einstein integration, see below. 
Subsequently, we determine the triple point as the intersection of the fluid-crystal and gas-crystal binodals, which occurs at $\beta p \sigma^3 = 0.0295$ and $\phi=0.59$. 
Finally, the integration of the gas-liquid binodal starts at the triple point.



We note that the binodals obtained from Gibbs-Duhem integration can be used to determine the chemical potential $\mu(p)$ as a function of pressure. For a given polymer reservoir packing fraction $\phi$, the Gibbs-Duhem binodals provide the coexistence pressures $p_{\rm coex,gx}$ and $p_{\rm coex,gl}$ for gas-crystal and gas-liquid coexistences, respectively. 
At these coexistence pressures, the chemical potentials of the coexisting phases are equal.
Therefore, starting from these pressures, we can easily integrate the equation of state $p(\rho)$ to obtain the (relative) chemical potentials of the gas, crystal, and liquid phases.
By inverting $\mu(p)$, we can also determine the pressure difference $\Delta p(\mu)$ that we need for our seeding simulations.

\subsection{Gas-crystal coexistence}
To determine an initial point for the integration of the gas-crystal binodal, we start by measuring the chemical potential of the face-centered cubic (fcc) crystal phase at zero pressure and a polymer reservoir packing fraction $\phi=0.8$ using non-equilibrium Einstein integration.\cite{freitas_nonequilibrium_2016} From this chemical potential $\mu(0)$ at zero pressure, we can accurately approximate the chemical potential of the crystal phase as 
$\mu_x (p) = \mu (0) + p / \varrho_0 $
where $\varrho_0$ is the number density of the crystal phase at zero pressure. 
This approximation follows from \Cref{eq:dmudp} and the approximation that the crystal density $\varrho$ is constant and equal to $\varrho_0$ in the considered pressure range.
For the gas phase, we can accurately approximate the chemical potential using the ideal gas law
$\beta \mu_g (p) = \log (\beta p \sigma^3)$. In both cases, the chemical potentials are defined relative to the chemical potential of an ideal gas at pressure $\beta p \sigma^3 = 1$.
By equating the chemical potentials of the gas and crystal phases, we find the gas-crystal coexistence pressure to be $\beta p \sigma^3 = 7.0 \times 10^{-4}$ at $\phi=0.8$.

\subsection{Spinodals from the equation of state}
To determine the spinodals of the gas and liquid phases, we used polynomial fits of the equations of state.
To be specific, for each phase, we first measured the equation of state $p(\varrho)$ using molecular dynamics simulations in the $NVT$ ensemble with $N=4\times 10^3$ particles. We then fitted the equation of state in the metastable region between the spinodal and binodal using a simple quadratic polynomial. From this fit, we calculate the colloid number density $\varrho$ at which the derivative $dp / d\varrho$ vanishes. Since this is equivalent to the point where $d^2G / d\varrho^2 = 0$, this criterion identifies the spinodal point.

\subsection{Scaling law fits of the binodals and spinodals}
Each dot in \Cref{fig:eq-phase-diagram} represents data obtained as  described above from the Gibbs-Duhem integration (binodals) or from the equations of state (spinodals). The lines in \Cref{fig:eq-phase-diagram} are fits and extrapolations of these data points. To determine the critical point, we fit the gas-liquid binodals as follows
\begin{align*}
    \eta_l(\phi) + \eta_g(\phi)  &= 2\eta_c + A \left(\phi^{-1} - \phi_c^{-1} \right), \\
    \eta_l(\phi) - \eta_g(\phi) &= B \left (\phi^{-1}  - \phi_c^{-1} \right)^\beta,
\end{align*}
i.e.\ the sum and difference of the liquid branch $\eta_l(\phi)$ and gas branch $\eta_g(\phi)$  are fitted linearly and to a scaling law with an exponent $\beta=0.32$, respectively.\cite{frenkel_understanding_2001} The fitting  parameters include the proportionality constants $A$ and $B$ as well as the colloid and polymer reservoir packing fractions $(\eta_c, \phi_c)$ at the critical point. However, away from the critical point, particularly for $\phi\gtrapprox 0.65$, we find that the gas-liquid  binodals and spinodals are less accurately described by these scaling laws. Therefore, to fit and extrapolate all binodals and spinodals in this region, we employ cubic spline fits.
The gas branches are generally fitted in log-space, i.e.\ $\log(\eta_g)$ is fitted as a cubic spline.
For the liquid phase, it is difficult to equilibrate for $\phi>0.8$ as the liquid crystallizes rapidly. Therefore, we rely on extrapolations of the cubic splines for the gas-liquid binodals in this regime. The extrapolations are simple linear extrapolations based on the gradient of the cubic splines at $\phi=0.8$.

\section{Determination of the kinetic phase diagram}
\label{sec:seeding-details}
\subsection{Preparation of seeds}
We prepare liquid and crystal seeds of $15$ different sizes, ranging from   $100$ to $5 \times 10^3$ particles. While we only use seeds with at least $500$ particles to calculate the nucleation barriers, smaller seeds are used as initial configurations for the transition path sampling simulations. These seeds are prepared separately for each polymer reservoir packing fraction $\phi$ between $\phi=0.6$ and $\phi=1.0$ in steps of $0.05$, resulting in approximately  $150$ different seeds in total.
To create a seed of a given size $n^*$, we equilibrate the `child' phase (liquid or crystal) at zero pressure and cut a spherical region from the bulk phase containing around $n^*$ particles. We then insert the seed into a low-density gas phase such that the volume of the gas phase is twenty times larger than that of the seed. The seed is equilibrated in the $NVT$ ensemble for $100,000$ timesteps. This procedure often leads to spontaneous crystallization of the liquid seed for $\phi\geq 0.85$, or spontaneous melting of a crystal seed for $\phi\leq 0.75$. These `transformed' seeds are not considered when computing the interfacial tensions $\gamma$ or the nucleation barriers $\Delta \Omega^*$. Consequently, the liquid-gas interfacial tension is based on seeds with $\phi\leq 0.85$, while  the crystal-gas interfacial tension is determined from seeds with $\phi\geq 0.75$.

\subsection{Nucleus size metric}
Inspired by previous work,\cite{zimmermann_nacl_2018} we employ  the following criterion to measure the size $n^*$ of a seed.
We first identify for each particle the number of nearest neighbors using the solid-angle-based nearest neighbor algorithm \cite{van_meel_parameter-free_2012} with anisotropy correction (ASANN).\cite{staub_parameter-free_2020}
We additionally require the cutoff radius determined by ASANN to be at most $1.6\sigma$.
If a nucleus is crystalline, it is identified as the largest cluster of particles with at least six nearest neighbors. This nearest-neighbor threshold of six is midway between the average number of nearest neighbors in the gas phase (zero) and the fcc crystal phase (twelve). 
If a nucleus is liquid, we use a nearest-neighbor threshold of five, which is midway between the gas phase (zero) and the liquid phase (ten).
In the analysis of the transition path sampling simulations (\Cref{fig:tps-cnt-9}), we use a nearest-neighbor threshold of five, irrespective of crystallinity.
Visual inspection of the crystal seeds (\Cref{fig:seeding}) reveals  that they are faceted. This threshold accounts for particles located on the faces, edges, and corners of these facets.
In the example nuclei depicted in \Cref{fig:seeding}, only the particles shown with the lightest beige color are excluded by the nearest-neighbor threshold.

\subsection{Critical conditions from active learning}
The critical chemical potential $\mu^*$, which corresponds to the chemical potential at which a given seed becomes critical, is determined using the active learning procedure described in Ref.\ \citenum{gispen_kinetic_2022}.
We will summarize this method here  with appropriate modifications.
This procedure essentially employs a bisection method with logistic regression taking care of the inherent stochasticity.
For each seed, we first determine a reasonable interval for the critical chemical potential, typically between the binodal and spinodal.
Subsequently, we repeat the following three steps:
\begin{enumerate}
    \item Estimate the critical chemical potential, denoted as  $\mu^*_{\textrm{\small est}}$, using logistic regression.
    \item Simulate the seed at the estimated chemical potential $\mu^*_{\textrm{\small est}}$ for at most $10^7$ timesteps, or until it grows to twice its size or shrinks to a nucleus of fewer than $10$ particles.
    \item Add ($\mu^*_{\textrm{\small est}}$,$B$) to the list of observations, where $B\in\{0,1\}$ indicates whether a seed grew or melted. 
\end{enumerate}
In the first step, the estimated critical chemical potential  is defined as the chemical potential at which the regression predicts a $50\%$ probability of growth. Simulating at this estimated critical chemical potential corresponds to the so-called `uncertainty sampling' approach to active learning. \cite{settles_active_2009} 
The use of logistic regression  circumvents the need to estimate the growth probability at a single chemical potential through multiple runs.
The final estimate for the critical chemical potential is obtained after fitting $20$ observations obtained in the way just described.
While this may seem a small amount of observations to approximate the critical chemical potential $\mu^*$, we find that the growth probability of a seed typically increases sharply from $0$ to $1$ within a small chemical potential range.
Therefore, the active learning procedure described above quickly finds a reasonable estimate of the critical chemical potential.
Instead of using hundreds of simulations to pinpoint exactly the conditions of $50\%$ growth for a single seed, this procedure allows to efficiently approximate the critical conditions of many crystal and liquid seeds for different polymer reservoir packing fractions.

\section{Transition path sampling}
\label{sec:tps}


In this appendix, we provide the details of our transition path sampling simulations. While the methodology closely follows  that recently described in Ref. \citenum{gispen_bcc_2025},  we repeat the explanation here with necessary modifications to ensure  that this paper is self-contained. Simulation management is done with the pyretis software package.\cite{riccardi_pyretis_2020}

\subsection{Aimless shooting} 
We employ the aimless shooting variant\cite{peters_obtaining_2006, mullen_easy_2015} of transition path sampling. In each shooting move, a ``shooting point'' (a time slice) is selected from the previous trajectory, where the velocities of all particles are completely resampled from the Maxwell-Boltzmann distribution. 
Starting from this new shooting point, we simulate forward and backward trajectories using grand-canonical Langevin dynamics.
A new transition path is accepted if one of the trajectories ends  in the metastable gas phase and the other ends  in the condensed phase. 
The trajectories can vary in  length, with the simulation stopping  when one of the stable states is reached.
To identify these stable states, we calculate the nucleus size $n$ every $10,000$ timesteps by identifying the nucleus as the largest cluster of particles that have  at least two neighboring particles within a distance of $r/\sigma < 1.6$.
We define the gas phase as $n<10$, while the condensed phase corresponds to $n>300$ for simulations starting with a liquid seed of $n^*=100$, $n>400$ for those starting with $n^*=200$, and $n>700$ for those starting with $n^*=500$.

\subsection{Shooting point selection}
The selection of shooting points follows a procedure slightly modified  from that described in Ref. \citenum{peters_obtaining_2006}. The initial shooting point is the critical liquid nucleus  obtained from the seeding simulations. For all subsequent shooting moves, the new shooting point $t_n$ is selected by slightly perturbing the previously accepted shooting point $t_o$. To be specific, $t_n$ is chosen randomly from the following candidates: $t_o, t_o \pm 1 \Delta T, t_o \pm 2 \Delta T, \dots, t_o \pm 20 \Delta T$, where $\Delta T = 10 \sqrt{\beta m \sigma^2}$, This $\Delta T$ should not be confused with the simulation timestep, $\Delta t = 0.001 \sqrt{\beta m \sigma^2}$.
The length of accepted trajectories varies with the state point. The shortest trajectories are seen for $\phi=0.8$ and $n^*=200$, where the average length is $120 \Delta T$. The longest trajectories are observed at $\phi=1.0$ and $n^*=500$, where the average length is $1050 \Delta T$. The time difference $|t_n - t_o|$ is typically much smaller than the total trajectory length. 

\subsection{Path decorrelation}
For each state point, $2000$ shooting moves were performed, yielding an average acceptance rate of $22\%$. The first $1000$ moves were used for equilibration, while the remaining $1000$ were used for production. We calculated the autocorrelation function of the nucleus size $n$ at the shooting point. To be more precise, we calculated $\mathbb{E} \left [(n_{k+l} - \bar{n})(n_{k} - \bar{n}) \right ]$, where $n_k$ is the nucleus size at the shooting point of shooting move $k$, $\bar{n}$ is the average nucleus size of shooting points, and $l$ is the shooting point lag. Based on this autocorrelation function, we estimated the decorrelation time to be between $10$ and $50$ shooting moves, depending on the specific state point. Consequently, we obtained between $20$ and $100$ decorrelated nucleation pathways,  depending on the state point.

\subsection{Reweighted transition path ensembles}
Recently, \citet{falkner_revisiting_2024} demonstrated that the aimless shooting move with flexible path lengths, as described above, is not reversible and introduces a bias into the transition path ensemble. To address this issue, they proposed a reweighting method where each transition path is assigned a relative weight of $1/L$, with $L$ representing the path length. To evaluate whether this issue affects our results, we applied their reweighting procedure to our transition path ensemble sampled with the aimless shooting point selection method explained above. Additionally, we repeated the transition path sampling simulations with a different shooting point selection method: instead of perturbing the previously accepted shooting point, we also performed a series of simulations where the shooting point was chosen uniformly and randomly from the previously accepted transition path. In this case, we used a modified acceptance criterion: a new transition path $X_n$ is accepted if one of the trajectories ends  in the metastable gas phase and the other ends in the condensed phase, with a probability $p_{\mathrm{acc}} = \min \left [1, L(X_o) / L(X_n) \right ])$.\cite{falkner_revisiting_2024} 
The acceptance rate with this procedure was significantly lower, ranging from approximately $5-10\%$ for most state points but dropping below $1\%$ for state point (a) because the trajectories at that state point are very long.
Thus, we obtained three different transition path ensembles. For each ensemble, we recalculated the average crystallinity $n_{\mathrm{crystal}} / n$, as a function of nucleus size $n$.

\begin{figure}[ht]
    \centering
    \includegraphics[width=\linewidth]{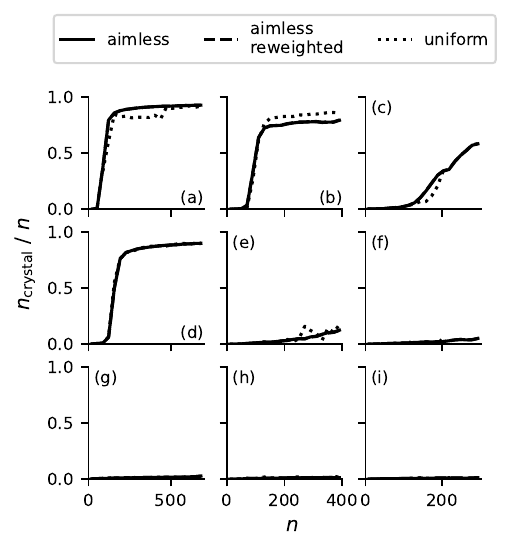}
    \caption{The average crystallinity $n_{\mathrm{crystal}} / n$ as a function of  nucleus size $n$ calculated for three different transition path ensembles: (1) the ensemble sampled with aimless shooting ("aimless"), (2)  the ensemble sampled with aimless shooting and reweighted by path length ("aimless reweighted"), and (3) the ensemble sampled using uniform shooting point selection ("uniform"). The letters refer to the state points indicated in the kinetic phase diagram \Cref{fig:kinetic-phase-diagram}.}
    \label{fig:tps-ensembles}
\end{figure}

In \Cref{fig:tps-ensembles}, we plot the crystallinity $n_{\mathrm{crystal}} / n$ for  three different transition path ensembles: (1) the ensemble sampled with aimless shooting ("aimless"), (2) the ensemble sampled with aimless shooting and reweighted by path length ("aimless reweighted"), and (3) the ensemble sampled using uniform shooting point selection ("uniform").
\Cref{fig:tps-ensembles} shows that the qualitative conclusions regarding the nucleation mechanisms are consistent across all path ensembles. 
The effect of reweighting seems minimal and is not discernible in \Cref{fig:tps-ensembles}.
The trajectories sampled using uniform shooting point selection seem to have slightly different crystallinities for state points (a,b,c,e).
Because the acceptance rate for uniform shooting point selection is significantly lower, it is possible that these differences are a result from slower convergence and/or a limited number of decorrelated pathways.
Overall, we conclude that the conclusions discussed in the main text are robust.




\section{Brute-force simulations of liquid droplet crystallization}
\label{app:bulk-vs-surface}

In the transition path sampling simulations presented in the main text, the liquid droplets are relatively small $N < 700$ when crystallization occurs. 
For this reason, it is hard to determine whether or not crystallization occurs at the surface of the droplet or whether it is facilitated by heterogeneous nucleation at the liquid-gas interface.
To  investigate this possibility further, we simulate large liquid droplets of $N=5\times 10^3$ colloids using brute-force molecular dynamics simulations in the canonical ($NVT$) ensemble. In this ensemble, the liquid droplets are metastable, showing no significant growth or shrinkage over time.  For polymer reservoir packing fractions $\phi\gtrapprox 0.77$, we observe spontaneous crystallization of these metastable droplets within a reasonable simulation time. We use polyhedral template matching (PTM)\cite{larsen_robust_2016} to identify crystalline particles as described in  \Cref{sec:ptm}. At the top of \Cref{fig:bulk-surface-rates}, we present a representative crystallization event of a liquid droplet at $\phi=0.78$. Liquidlike particles are depicted in dark blue and reduced in size, while crystalline particles are shown in red. We see that crystallization predominantly occurs in the bulk of the droplet rather than at the liquid-gas interface.

\subsection{Crystal nucleation rate in bulk liquids and liquid droplets}
To support this claim, we  measure the crystal nucleation rate using brute-force molecular dynamics simulations as a function of the polymer reservoir packing fraction $\phi$. For $\phi$ values between $0.77$ and $0.81$, we perform $16$ simulations each for a pure bulk liquid and a liquid droplet surrounded by a gas phase. The bulk liquid contains $N=4\times 10^3$ colloids, while the  liquid droplets contain $N=5\times 10^3$ colloids. The bulk liquid is simulated at zero pressure in the isobaric-isothermal ($NPT$) ensemble, whereas the liquid droplet is simulated in the canonical ($NVT$) ensemble. In this way, both systems are simulated under conditions close to  gas-liquid coexistence. 
We use the same methodology as in Ref.\ \citenum{gispen_brute-force_2023} to compute the nucleation rate and its $95\%$ confidence intervals from our simulations. The nucleation rate density $J$ is computed using the number of observed crystallization events $\ell$, the total simulation time $t$, and the liquid volume $V$ as
\begin{equation*}
    J = \ell / (V t).
\end{equation*}
For $\phi=0.77$, the nucleation rate is so low that only a few systems crystallized: $\ell=4$ for  bulk liquids and $\ell=7$ for  liquid droplets. In all other cases, $\ell=15$ or $16$,  i.e.\ almost all systems crystallized. For  liquid droplets, we approximated the liquid volume $V$ as the number of particles in the droplet divided by the liquid number density at coexistence conditions. 

\begin{figure}
    \centering
    \includegraphics[width=\linewidth]{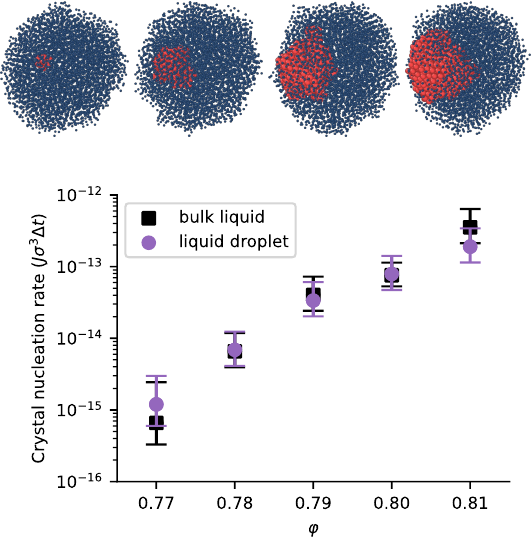}
    \caption{Bulk versus surface crystallization of liquid droplets. At the top, a representative crystallization event of a liquid droplet of $N=5\times 10^3$ colloids is shown. Liquidlike particles are depicted in dark blue and reduced in size, while crystalline particles are shown in red. At the bottom, we plot the crystal nucleation rate $J$ measured using brute-force simulations as a function of the polymer reservoir packing fraction $\phi$. 
    The nucleation rate $J$ is normalized by $\sigma^3$ with $\sigma$ the colloid diameter  and by the simulation timestep $\Delta t = 0.001 \sqrt{\beta m \sigma^2}$.
    The black squares represent the crystal nucleation rate measured in bulk liquids, while the purple circles show those measured in liquid droplets containing $N=5\times 10^3$ colloids. The error bars represent approximate $95\%$ confidence intervals for the nucleation rates.
    }
    \label{fig:bulk-surface-rates}
\end{figure}

In \Cref{fig:bulk-surface-rates}, we plot the crystal nucleation rate for bulk liquids (black squares) and liquid droplets (purple circles). 
The error bars represent $95\%$ confidence intervals and are at most approximately half an order of magnitude.
If  crystallization were a heterogeneous nucleation process facilitated by the liquid-gas interface, the crystal nucleation rate should be significantly higher for liquid droplets than for bulk liquids.
However, the crystal nucleation rate in bulk liquids is consistent with  that in liquid droplets within our error bars. 
Therefore, our nucleation rate measurements support our claim that  crystallization in droplets occurs in the bulk of the droplet rather than at the liquid-gas interface.


\begin{figure}
    \centering
    \includegraphics[width=\linewidth]{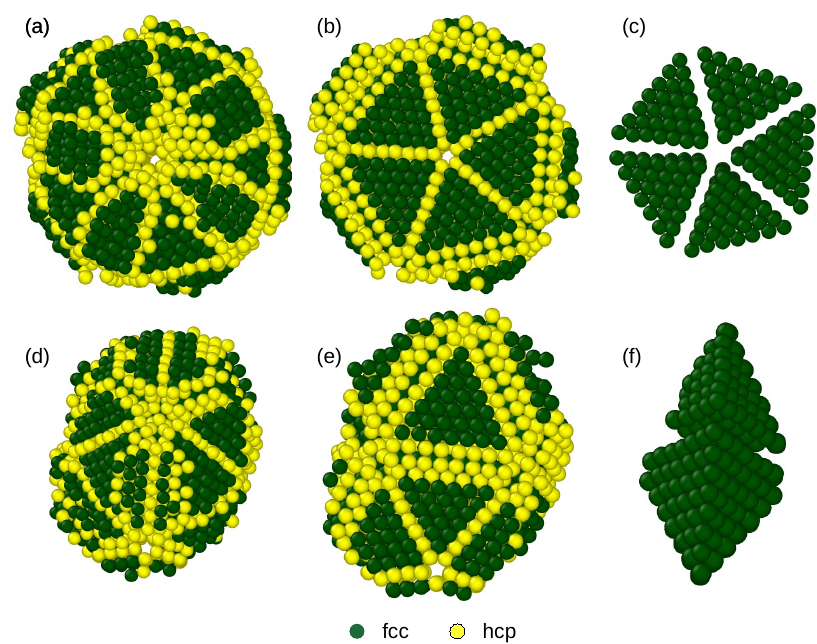}
    \caption{Fivefold symmetry in a spontaneously crystallized cluster of $N=5\times 10^3$ colloids. 
    The top row (a-c) and bottom row (d-f) show different perspectives of the same cluster. Panels (a,d) show the external surface, panels (b,e) show a cut through the center, and panels (c,f) show the central arrangement of five tetrahedra forming a pentagonal bipyramid.
    The local crystal structure is classified using the polyhedral template matching (PTM) algorithm. Dark green particles are fcc-like,  yellow particles  are hcp-like, and liquidlike particles are not shown.
    }
    \label{fig:fivefold}
\end{figure}

\subsection{Fivefold symmetry clusters}

When analyzing the brute-force crystallization events of  liquid droplets containing $N=5\times 10^3$ colloids, we find that the crystallized clusters show a peculiar fivefold symmetry. 
This fivefold symmetry is consistently seen  in  clusters formed at polymer reservoir packing fractions $\phi=0.77, 0.78, 0.79$. However,  for $\phi=0.8$, the clusters show a random hexagonal close-packed (rhcp) structure instead.
In \Cref{fig:fivefold}, we show a crystal cluster  formed at  $\phi=0.78$.
The figure presents  visualizations of the cluster from two orthogonal perspectives in different rows. 
The particles are colored according to their local crystal structure, as classified using polyhedral template matching (PTM),\cite{larsen_robust_2016} see \Cref{sec:ptm} for details.
Dark green particles are fcc-like, while  yellow particles are hcp-like.
Panels (a,d) show the external surface of the cluster, while panels  (b,e) provide a cross-sectional view, where we slice the cluster in half to show its internal structure. Finally, panels (c,f)  show only fcc-like particles in the core of the cluster.


The core of the cluster (c,f) is an arrangement of five tetrahedral-shaped domains of fcc-like particles. Together, these tetrahedral domains form a pentagonal bipyramid. 
Hcp layers are observed between the tetrahedral domains and as additional surface layers (b,e).
The structure of this cluster is reminiscent of the clusters observed for hard spheres self-assembling under spherical confinement.\cite{de_nijs_entropy-driven_2015}
In spherical confinement, hard spheres can self-assemble into both icosahedral clusters and decahedral clusters.\cite{de_nijs_entropy-driven_2015,mbah_early-stage_2023} Although Refs. \citenum{de_nijs_entropy-driven_2015},\citenum{mbah_early-stage_2023} observe a preference for icosahedral clusters, we only observe decahedral clusters like the one shown in \Cref{fig:fivefold}. 
Our simulations differ from those of  spherically-confined hard spheres in two ways. First, we consider polymer-induced attractions between the hard spheres. Second, the spherical confinement in our system arises from the polymer-induced attractions between the colloids, rather than  from a hard external wall.
As a result, the spherical confinement in our system is much `softer', which also helps explain the absence of  heterogeneous nucleation in our simulations.
Our results are also reminiscent of fivefold-twinned crystals observed in various systems, such as nucleation from  bulk hard-sphere systems,\cite{omalley_crystal_2003} gold nanoparticles,\cite{koraychy_growth_2022} and many other metals.\cite{hofmeister_forty_1998} 
In fact, in our bulk liquid simulations, we observe fivefold-twinned crystals that closely resemble  those found in Ref.\ \citenum{omalley_crystal_2003}.
Our results suggest that the fivefold symmetry can be further stabilized by polymer-induced spherical confinement, which  could serve as an interesting starting point for future research.


\section{Polyhedral template matching}
\label{sec:ptm}
Polyhedral template matching (PTM) \cite{larsen_robust_2016}  identifies local crystal structures by matching the positions  of a set of  particles  to  reference templates corresponding to various crystal phases.
For face-centered cubic (fcc) and hexagonal close-packed (hcp) structures, the template consists of thirteen particles: a central particle and its twelve nearest neighbors, arranged according to  the fcc or hcp crystal lattice. 
We use a root-mean-square deviation (RMSD)  threshold of $0.12$ for the PTM algorithm.
In \Cref{fig:fivefold}, we identify the central particle of a recognized set as either fcc-like or hcp-like.
For the main text, we define a particle as `crystalline' if it is part of one of the thirteen particles in a recognized fcc or hcp set.
By classifying all thirteen particles in the set as crystalline, we also include particles located on the crystal surface  as crystalline. 


%

\end{document}